\documentclass[]{aastex631}

\begin{document}

\title{The Double-White-Dwarf Merger Rate from ZTF }

\author[0000-0002-6579-0483]{Dan Maoz}
\affiliation{School of Physics and Astronomy \\
Tel Aviv University \\
Tel Aviv 69978, Israel}



\begin{abstract}

Using the Zwicky Transient Facility (ZTF), Burdge et al. (2020) discovered systems of eclipsing double white dwarfs (EDWDs) having orbital periods $<1$~hr. From the properties of 3 of the discovered systems, I estimate a merger rate of DWDs, 
per unit stellar mass in the Galaxy, of $R_{\rm merge, M*}\approx4.8\times 10^{-13}$~yr$^{-1}$~M$_\odot^{-1}$.
 This likely somewhat underestimates the rate, because of several known
 effects that work against EDWD detection in ZTF. The derived merger rate is within the uncertainty range, $R_{\rm merge, M*} =(4.6-5.8)\times 10^{-13}$~yr$^{-1}$~M$_\odot^{-1}$, measured independently by Maoz et al. (2018) based on two samples of DWDs discovered via radial-velocity variations. Based on the expected period distribution of DWDs and their detectability, of order 100 additional eclipsing DWDs with  periods $>1$~hr are discoverable in ZTF, with potential to significantly improve the merger-rate's measurement precision.

\end{abstract}



\section{Introduction} \label{sec:intro}

Binary systems consisting of two white dwarfs (WDs) at close separations lose energy and angular momentum to the emission of gravitational waves, leading to their in-spiral and eventual mergers.
The rate at which such mergers occur in the Milky Way and similar galaxies is of great interest for a number of reasons. For example, the leading contending scenario for the progenitor event of Type Ia supernovae (SNe Ia; see e.g. \citealt{Maoz2014}), is the merger of some exemplars of these double WDs (DWDs; of precisely which exemplars is yet unclear). 
Comparison of the DWD merger rate to the SN Ia rate is a direct test of this scenario. 
As another example, in the planned LISA gravitational wave observatory, short-orbit DWDs will constitute the most common type of resolved detected sources, but also the main source of noise that sets the detection floor of the experiment, in the form of the combined emission from numerous unresolved DWDs (e.g. \citet{Korol2022}). Knowing the numbers, masses, and period distribution of the DWD population (which together dictate their merger rate) is necessary in preparation for the mission.

Some of the properties of the DWD population and its merger rate have been estimated by searching WD samples for radial-velocity variations (RVV) between two or more epochs, indicative that a WD has a close binary companion that is also a WD (as sometimes confirmed by a double-lined spectrum, although generally not). \cite{Maoz2012} and \cite{BadenesMaoz2012} used this method to discover 15  candidate DWD
 systems among the spectra of 4000 WDs from the Sloan Digital Sky Survey (SDSS). \cite{MaozHallakoun2017} found 43 DWD candidates among 439 WDs from the ESO-SPY survey \citep{Napiwotzki2020}. \cite{Maoz2018} analyzed jointly the results of the two samples and derived a DWD merger rate per unit stellar mass in the Galaxy of $R_{\rm M*} =(4.6-5.8)\times 10^{-13}$~yr$^{-1}$~M$_\odot^{-1}$.
 This DWD merger rate is about 6 times the SN Ia rate in Milky-Way-like galaxies, indicating that, if a large enough fraction, $\gtrsim 15\%$, of all DWD mergers lead to a SN Ia explosion (rather than to some other result, say, to a merged massive WD), then DWD mergers are a viable SN Ia progenitor scenario.

 \cite{Burdge2020} have performed a search for eclipsing DWDs with orbital periods up to 1~hr in photometric data from the Zwicky Transient Facility (ZTF). In this paper, I analyze their reported results and use then to derive a new, independent, estimate of the DWD merger rate. I then estimate the number of eclipsing DWDs with longer periods that are still lurking, but discoverable, in the ZTF data.  

\section{A ZTF-based white-dwarf merger rate} \label{sec:style}
In the main part of the survey for DWDs reported by \cite{Burdge2020}, the ZTF light curves of 10 million blue stars, selected based on Pan-STARRS color criteria, and having over 50 ZTF epochs each, were searched for periodic behavior in the period range 6-60~min, using an assortment of search algorithms. \cite{Burdge2020} thus discovered  15 DWDs,
7 of them eclipsing DWDs (EDWDs). However the chosen color-based parent-sample selection was broad and inclusive of 
many types of objects, overwhelmingly {\it not} WDs. It is thus difficult to estimate the actual number of WDs that were searched for periodicity, a number necessary for deriving a merger rate. Most of the 15 discovered DWDS are blue and overluminous compared to typical WDs, i.e. this sample is strongly biased toward
large-radius (i.e. low-mass), hot (i.e. young) WDs or pre-WDs, often stripped by previous interactions between the companions (which was indeed the stated motivation of \cite{Burdge2020} for focusing on a blue parent population). 

Fortunately, \cite{Burdge2020} describe in a paragraph in their section 5.1 how they repeated their search for periodicities, but in another parent sample: 248,000 ZTF light curves with $>50$ epochs, of sources that are candidate WDs in the {\it Gaia}-based analysis of \cite{GentileFusillo2019}. According to the latter, only about half of these WD candidates are in fact high-confidence ($P_{\rm WD}>0.75$) WDs, the rest of the sample having varying levels of contamination by non-WDs. I will assume, therefore, that 3/4 of the parent sample, i.e. 186,000 of these ZTF light curves, are of true WDs.  The \cite{Burdge2020} analysis of this sample recovered three of the above seven EDWDs, and no additional EDWDs. These 3 non-accreting  EDWD systems (ZTFJ0538+1953, ZTFJ0722-1839, and ZTFJ1901+5309) have 
orbital periods of 14~min, 24~min, and 41~min, respectively, are are quite similar: all but one of the six WD masses are around 0.35~M$_\odot$ (one WD has 0.45~M$_\odot$) ; all but one of the six WD radii are about 0.023~R$_\odot$ (one WD has 0.029~R$_\odot$). 

However, it appears that \cite{Burdge2020} detected only a fraction of all of the EDWDs among the \cite{GentileFusillo2019} sample that they analyzed.   All three of the detected DWDs are brighter than $\sim19$~mag, even though the \cite{GentileFusillo2019} sample extends to 20~mag and beyond. Indeed, \cite{Burdge2020} mention that ZTF photometry having $S/N<5$ is completely omitted from the ZTF database. As a result,
in fainter targets, the photometric measurements within the eclipse, when the target dims, may be absent, and hence the eclipse and the system might completely evade detection.
I estimate that only about 40\% of the \cite{GentileFusillo2019} high-confidence WDs are brighter than 19~mag. If, in fact, an un-eclipsed magnitude of 19 mag is a requirement for DWD eclipse detection in the \cite{Burdge2020} analysis, then their WD parent sample in which eclipses are detectable was effectively of only $N_{\rm parent}\approx75,000$ Gaia WDs.
The three EDWD's Gaia absolute magnitudes, $M_G$, are  2-3~mag brighter than typical WDs with similarly high temperatures.  Large and low-mass  WDs such as in these three systems have double the eclipse probability of the typical (cool, $\sim0.6$~M$_\odot$) WD in the
\cite{GentileFusillo2019} Gaia sample. All three also have over 500 ZTF epochs. All of these elements may have affected the detection efficiency in this search.

I estimate the DWD merger rate as follows.
For a random inclination of the orbital planes, the eclipse probability by a random observer is $P_{\rm e}=(R_1+R_2)/a=0.26, 0.19$, and 0.15, for each of the three EDWDs, respectively, given their WD radii $R$ and orbital separations $a$ .
In other words, the 3 discovered EDWD systems are representative of 
\begin{equation}
N_{\rm dwd}=\displaystyle\sum_{i=1}^3 P_{\rm e,i}^{-1}=15.9
\end{equation}
similar DWD systems of all inclinations in the sample. For the EDWD with the widest orbit of the three, 41~min, the gravitational-wave-driven merger time (which depends primarily on the separation, as $a^4$)  is $t_{\rm gw, max}\approx24$~Myr. The DWD merger rate, per WD in the Galaxy, is thus 
\begin{equation}
R_{\rm merge, WD}=\frac{N_{\rm dwd}}{t_{\rm gw, max} N_{\rm parent}}
\approx \frac{15.9}{2.4\times10^7{\rm yr}\times 7.5\times10^4}
=8.8\times 10^{-12}~{\rm yr}^{-1}.
\end{equation}
The dominant uncertainty in this rate, $\sim \pm60\%$, is the Poisson error in the sole 3 detections. For comparison to SN~Ia rates, it is useful to also express the DWD merger rate per unit stellar mass in the Galaxy, dividing by the ratio of Galactic stellar mass to WD number, $18.5 ~M_\odot$ per WD (see \cite{Maoz2018}), or
\begin{equation}
R_{\rm merge, M*}\approx4.8\times 10^{-13}~{\rm yr}^{-1}~{\rm M}_\odot^{-1}. 
\end{equation}

This ZTF-based DWD merger rate is in excellent agreement with the DWD merger rate derived by  \citet{Maoz2018} jointly from two RVV-based DWD searches,  $R_{\rm merge, M*} =(4.6-5.8)\times 10^{-13}$~yr$^{-1}$~M$_\odot^{-1}$.  
 This, without correcting for the further effect of the reduced eclipse probability (by up to factor 2) of more typical, higher mass and hence smaller WDs than those detected; nor for the possibility that EDWD detection in ZTF might in fact require $\gtrsim 500$ data epochs (it is unclear what fraction of the WDs had as many epochs in the \cite{Burdge2020} sample). In summary, the correct ZTF-based rate is likely somewhat even higher than the above estimate, confirming the   \citet{Maoz2018} conclusion that the merger rate of DWDs is a factor $\sim 6$ higher than the SN Ia rate, and perhaps even higher.

 The dominant error in my estimate, above, of the DWD merger rate, as noted, is the small number (3) of detected systems. However, this could be ameliorated by discovering, in ZTF, EDWDs with periods $>1$~hr. Although such DWDs merge on longer timescales (even greater than a Hubble time for periods above about 8~hr), that is irrelevant here---those longer-period DWDs are just members of another segment of the same Galactic DWD period distribution, whose shape and normalization sets the {\it future} merger rate. Going from separations $a=0.2$~R$_\odot$ (periods of $\sim 20$~min) to $a=2$~R$_\odot$ (periods up to $\sim10$~hr), the separation distribution of DWDs, $dN/d(\log a)$, grows by about 2.5 decades \citep{Maoz2012}.
 Thus there should be about 300 times more DWDs in the 1-10~hr period range than in the 0.1-1~hr range searched by  \cite{Burdge2020}. On the other hand, the eclipse probability decreases by a factor 10 at the larger separations, and therefore we might expect only $\sim30$ times more discoveries, i.e. of order 100 DWDs, in the longer-period range. This prediction could be tempered somewhat if, for observational reasons, 
 longer-period eclipsing systems in ZTF are significantly harder to find---there will be fewer eclipses per light curve, and fewer points (or even no points) per eclipse, and hence less-significant eclipse detections.
 Nevertheless, a ZTF search for EDWDs at longer periods is likely to increase by an order of magnitude the number of DWD systems that are useful for determining the DWD merger rate, improving its measured precision.   

\begin{acknowledgments}
This paper is part of a project that has received funding from the European Research Council (ERC) under the European Union’s Seventh Framework Programme, grant agreement No. 833031 (PI: Dan Maoz).
\end{acknowledgments}

\bibliography{sample631}{}
\bibliographystyle{aasjournal}



\end{document}